\documentclass[useAMS,usenatbib]{mn2e}
\usepackage{graphicx,psfig,cite,epsf}  
\usepackage{times}
\usepackage{subfigure}
\usepackage{textcomp}
\usepackage{multirow}
\usepackage{amsmath}
\usepackage{amsmath}	% Advanced maths commands
\usepackage{amssymb}	% Extra maths symbols
\usepackage{gensymb}

\def\apjl{ApJ}                % Astrophysical Journal, Letters
\def\apjs{ApJS}               % Astrophysical Journal, Supplement
\def\ao{Appl.~Opt.}           % Applied Optics
\def\aap{A\&A}                % Astronomy and Astrophysics
\def\src {1E 1547.0--5408}
\def\xmm{{\it XMM-Newton}}
\def\bat{{\it Swift/BAT}}

\title[X-ray dust-scattering of a 1E 1547.0--5408 burst]{The effect of X-ray dust-scattering on a bright burst from the magnetar 1E 1547.0--5408}

\author[Pintore F. et al.]{Fabio Pintore$^{1}$\thanks{E-mail:
pintore@iasf-milano.inaf.it},  Sandro Mereghetti$^{1}$,  Andrea Tiengo$^{2,1,3}$,   Giacomo Vianello$^{4}$, 
  \newauthor  
Elisa Costantini$^{5}$,  Paolo Esposito$^{6}$ \\
    $^1$ INAF -- IASF Milano, Via E. Bassini 15, 20133 Milano, Italy\\
   $^2$ Scuola Universitaria Superiore IUSS Pavia, Piazza della Vittoria 15, 27100 Pavia, Italy\\
  $^3$ Istituto Nazionale di Fisica Nucleare, Sezione di Pavia, Via A. Bassi 6, 27100 Pavia, Italy \\
   $^4$  SLAC National Accelerator Laboratory, Stanford University, Stanford, CA 94305, USA \\
     $^5$  SRON Netherlands Institute for Space Research, Sorbonnelaan, 2, 3584-CA, Utrecht, The Netherlands   \\
      $^6$ Anton Pannekoek Institute for Astronomy, University of Amsterdam, Postbus 94249, 1090-GE Amsterdam, The Netherlands            }

\date{Accepted  . Received  ;}

\pubyear{2017}

\begin{document}

%\label{firstpage}
\maketitle

\begin{abstract}
A bright burst, followed by an X-ray tail lasting   $\sim$10 ks, was detected during an \xmm\ observation of the magnetar   \src\ carried out on 2009 February 3. The burst,  also observed by  \bat ,  had a spectrum well fit by the sum of two blackbodies with temperatures of $\sim$4 keV and 10 keV and a fluence in the 0.3--150 keV energy range of $\sim10^{-5}$ erg cm$^{-2}$.  The X-ray tail had  a fluence of $\sim4\times10^{-8}$ erg cm$^{-2}$. Thanks to the knowledge of the   distances and relative optical depths of three dust clouds between us and \src , we show that most of the X-rays in the tail  can be explained by dust scattering of the burst emission,  except for the first $\sim$20--30 s.
We point out that  other X-ray tails observed after strong magnetar bursts may contain a non-negligible  contribution due to dust scattering.

\end{abstract}


\section{Introduction}

Dust grains in the interstellar medium cause small-angle scattering of soft X-ray photons.  
Due to this effect,  if there is a significant amount of interstellar dust along the line of sight, bright point-like X-ray sources may appear surrounded by diffuse X-rays (the so-called ``scattering halo''),  
as predicted by \citet{overbeck65} and first observed by \citet{rolf83} and \citet{catura83}. The study of such X-ray halos can provide important information on the properties of the interstellar dust     \citep[e.g.][]{mathis91,draine03,costantini05}.  

Due to their longer path length, scattered X-rays  have a time delay with respect to the unscattered  ones. In the case of   variable sources, this effect can be used to derive information on  the source distance and the spatial  distribution of the dust \citep{trumper73,predehl00,mir99}.
Short duration  bursts scattered by  thin layers of dust produce  dust-scattering rings which appear to expand with time around the central source.  Such rings were seen around galactic  binary systems \citep[e.g.][]{heinz15, heinz16, vasilopoulos16},    magnetars  \citep[e.g.][]{tiengo10, svirski11}, and  gamma-ray bursts \citep[e.g.][]{vaughan04,tiengo06,vianello07}.

The presence of delayed scattered radiation can also affect the time profile and spectrum of the X-rays detected after bright bursts. This must be taken into account when instruments with inadequate angular resolution, which do not permit to disentangle  the scattered and unscattered components, are used. For example, it has been suggested that some of the  X-ray afterglows of gamma-ray bursts, in particular those showing a long plateau phase, are  due to scattering from dust in the host galaxies \citep{shao07,shao08}. Although \citet{shen09}  showed that the lack of spectral evolution of most   afterglows is inconsistent with this hypothesis, a few cases in which at least part of the X-ray emission can be explained  by   dust scattering have been recently reported 
\citep[see, e.g.,][]{hol10,eva14short,wang16}.

In this work, we  evaluate  the effects of dust scattering on the observed properties of the X-ray tails seen after magnetar bursts,  using, as a case study, an \xmm\ observation of \src\ in which a burst followed by a decaying tail lasting $\sim$10 ks was detected. \src\ is a transient magnetar which showed three    outburst episodes \citep{bernardini2011} and is surrounded by extended X-ray emission  consistent with a  dust scattering halo     \citep{olausen11}. 
This source is  particularly suitable for our investigation since, thanks to previous analysis of three dust scattering rings caused by a much brighter burst, some of the properties of the dust in this direction are already known \citep{tiengo10}.

In Section~\ref{data_reduction} we present the data reduction of the  \xmm\  and  {\it Swift} observations; in Section~\ref{results} we first characterize the   properties of the persistent emission and of the burst of \src\ and then analyze the  time evolution of the radial profiles of the X-ray tail and we fit the tail light curve and spectrum with a dust scattering model.  We discuss our results in Section~\ref{discussion}, where we also compare  this event from \src\ with     other magnetar bursts and flares followed by extended tails.

\section{data reduction}
\label{data_reduction}

We analyzed an  {\it XMM-Newton} observation of \src ,   with exposure time of $\sim$56 ks, taken on 2009 February 3.  
This is the same observation used by \citet{tiengo10} to study the dust scattering rings produced by a bright burst that occurred on January 22 \citep{mereghetti09}. We note that the presence of such rings, at angular distance larger than 3$'$ from the central source, does not affect the analysis presented here.   

The three cameras of the EPIC instrument, one pn camera \citep{struder01short} and two MOS  cameras \citep{turner01short}, were operated in full-frame mode and with a thick optical blocking filter. 
We reduced the data with the  SAS v.14.0.0 software, selecting single- and double-pixel events ({\sc pattern}$\leq$4) for the pn and single-  and multiple-pixel events for the MOS ({\sc pattern}$\leq$12). Because of its high  count rate ($>6$ cts s$^{-1}$ in the EPIC-pn), the source data were affected by  pile-up, therefore 
we extracted the source spectra and lightcurves from an annular region with inner and outer radii of 5$''$ and 40$''$, respectively. The background was   extracted from a  circular region of radius  60$''$ free of sources. 
For the RGS instrument we obtained the source events following  the standard procedures described in the SAS threads\footnote{http://www.cosmos.esa.int/web/xmm-newton/sas-thread-rgs}.

We also used a \bat\ observation (Obs.ID: 00341965000) taken almost simultaneously with the \xmm\ one.  Spectra and lightcurves were extracted following the standard data reduction procedures described in the \bat\ threads\footnote{http://www.swift.ac.uk/analysis/bat/index.php}. 

The spectral fits were carried out with the   {\sc xspec v.12.8.2} software package,  adopting the {\sc phabs} model,   with the solar abundances  of   \citet{wilms00}, for the interstellar absorption. All the errors in the spectral parameters reported below are at the 90\% confidence level for a single interesting parameter.

\section{Data analysis and results}
\label{results}

\begin{figure}
		\hspace{-0.5cm}\includegraphics[width=6.cm,angle=270]{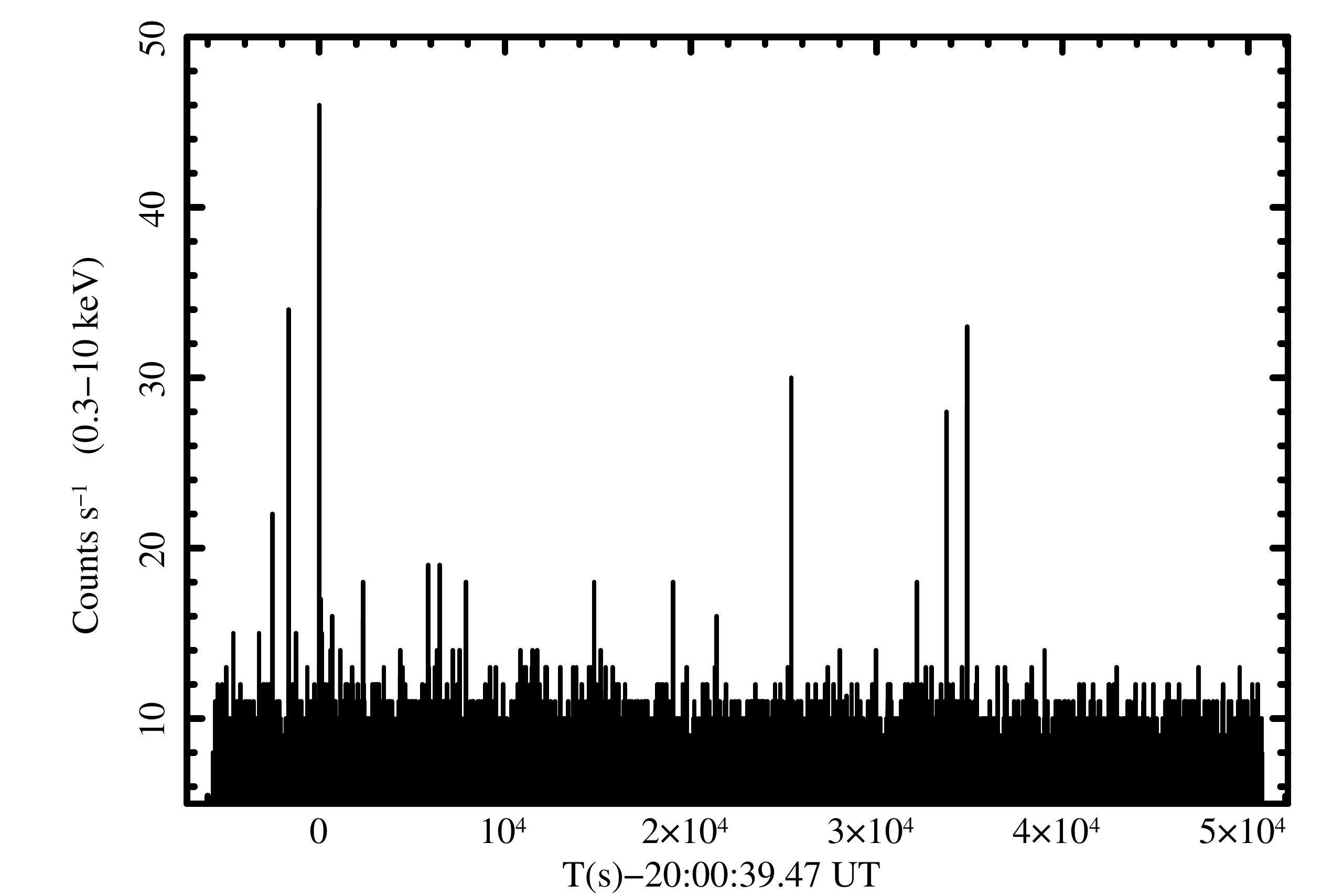}
   \caption{EPIC-pn  lightcurve of \src\  in the energy range 0.3--10 keV   binned  at 1 s. Several bursts are clearly visible. The strongest (and saturated) burst  at T=0 is the one with the X-ray tail considered in this work.}  
        \label{lc}
\end{figure}

\begin{figure}
	\hspace{-0.5cm}\includegraphics[width=6.0cm,angle=270]{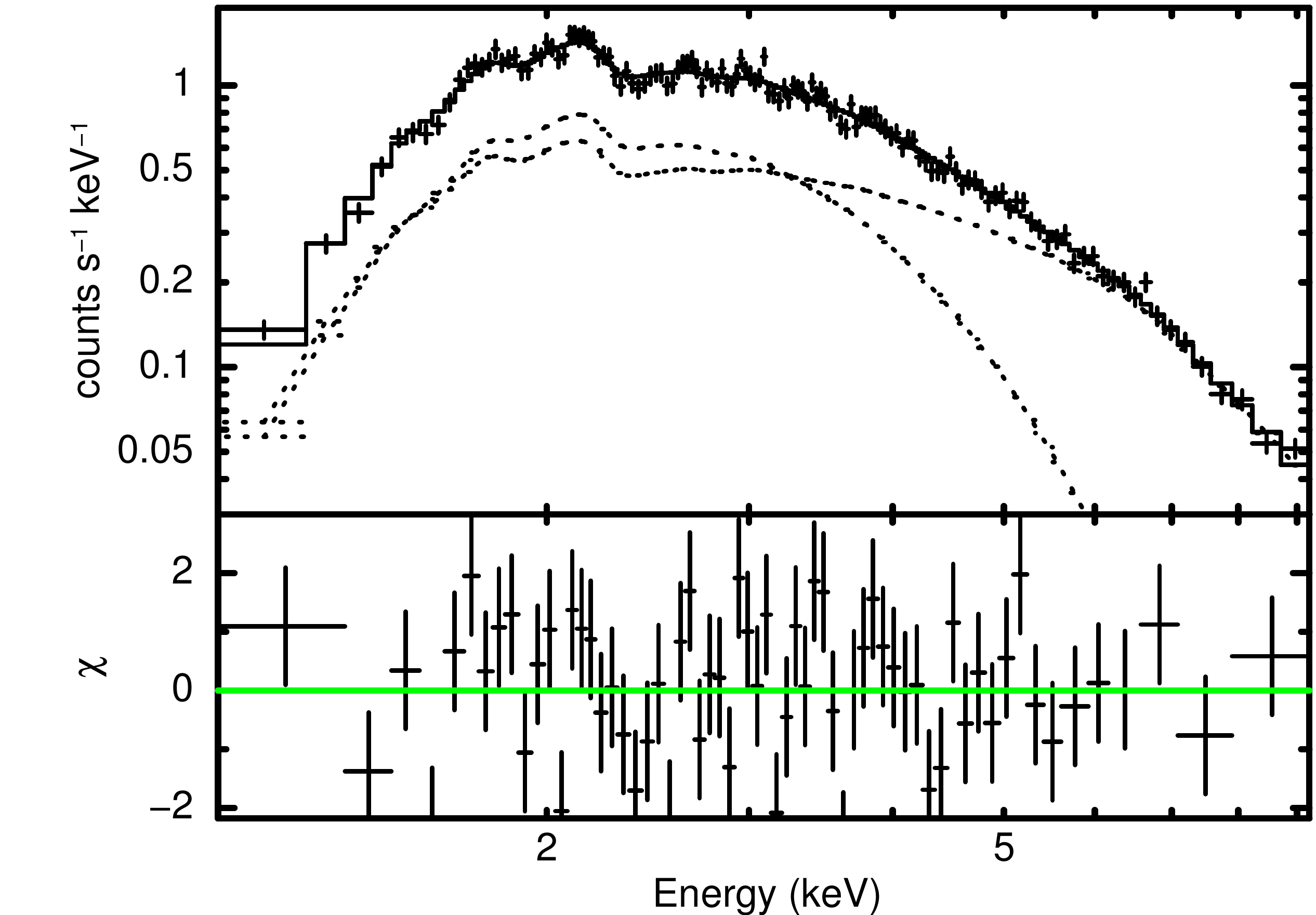}
    \caption{Top panel:   EPIC-pn spectrum of the persistent emission of \src\ derived from the last $\sim$9 ks of the \xmm\ observation. The solid line represents the best-fiting model, which consists of  a powerlaw (dashed) plus a  blackbody (point-dashed). Bottom panel: residuals of the best-fitting (rebinned for display purposes).}
        \label{pers}
\end{figure}

During the \xmm\ observation, \src\ emitted several short bursts (Fig.~\ref{lc}). In this work we concentrate on the brightest one, which was followed by a  tail of enhanced X-ray  emission lasting about 10 ks. 
In   the analysis of the EPIC data, we removed short time intervals corresponding to all the fainter bursts  visible during the observation.

We derived the properties of the persistent emission of \src\   from   the last $\sim$9 ks of the \xmm\ observation, when the source had the lowest count rate and no bursts were emitted. 
We extracted the  EPIC-pn spectrum, in the energy range 1--10 keV, from this time interval and fitted it with the phenomenological model usually adopted for magnetars in this energy range, i.e. an  absorbed  blackbody plus power-law model \citep{mereghetti08}. We obtained a good fit ($\chi^2/dof=139.55/128$, see Fig.~\ref{pers}) with the following  parameters: column density N$_{\rm H}$=$(4.2\pm0.3)\times10^{22}$ cm$^{-2}$,  blackbody  temperature kT=0.64$\pm$0.03 keV,  emitting radius $R$=$(1.8\pm0.2)\cdot d_{4\text{kpc}}$  km, power-law photon index   $\Gamma$=1.8$\pm$0.3 and normalization F$_{\rm 1 keV}$=$1_{-0.5}^{+0.8}\times10^{-2}$  photons cm$^{-2}$ s$^{-1}$ keV$^{-1}$.
The   absorbed and unabsorbed fluxes in the 1--10 keV energy range are $(4.59\pm0.06)\times10^{-11}$ erg cm$^{-2}$ s$^{-1}$  and $(8.3\pm0.2)\times10^{-11}$ erg cm$^{-2}$ s$^{-1}$, respectively.
These results are consistent with those reported   by \citet{bernardini11},  who did not exclude the time interval of the burst tail from their spectral analysis.
 
  \begin{figure}
	\subfigure{\includegraphics[height=6.1cm]{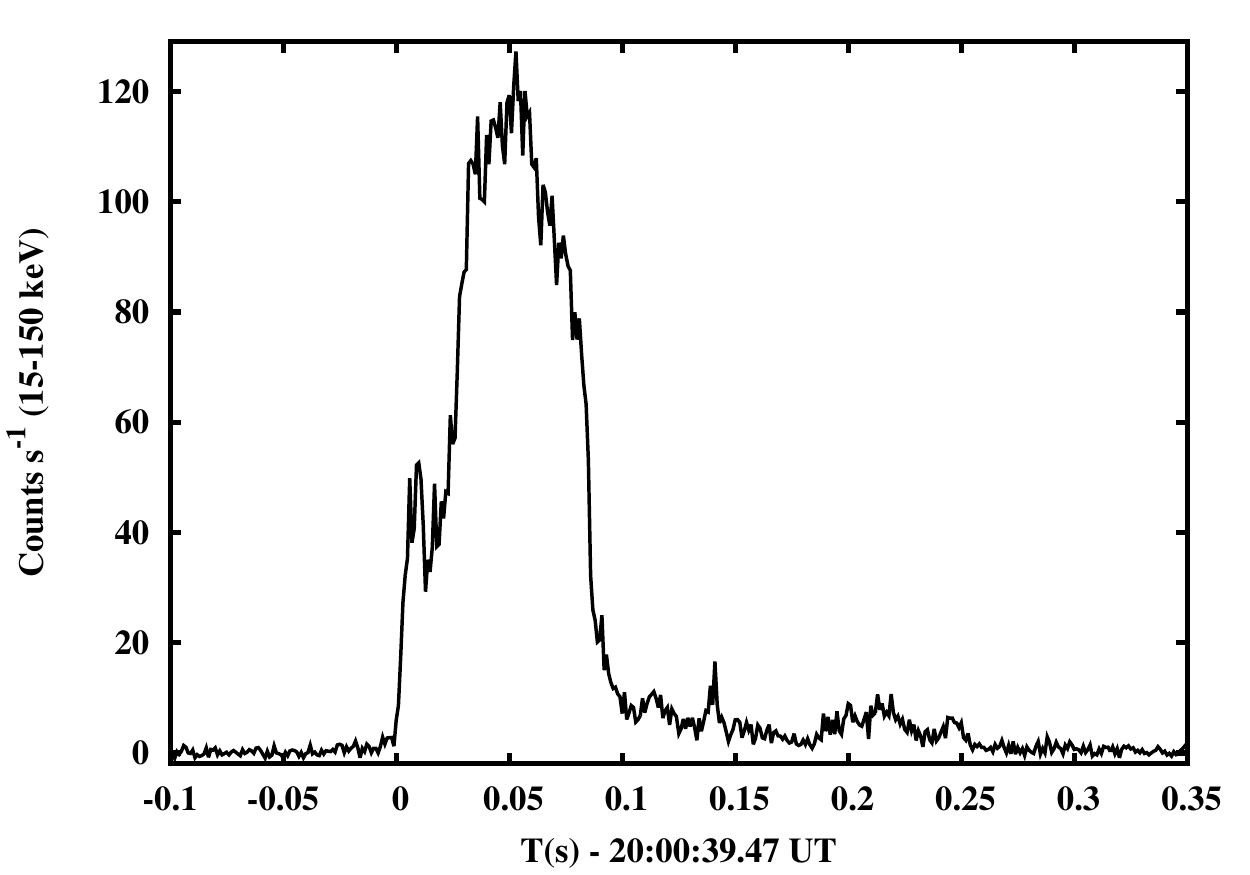}}
\vspace{-0.5cm}
    \caption{\bat\ lightcurve of  the bright burst    in the 15-150 keV energy range with a bin size of 1 ms.    }
        \label{bat_lc}
\end{figure}

 \begin{figure}
	\hspace{-0.5cm}\includegraphics[width=6.3cm,angle=270]{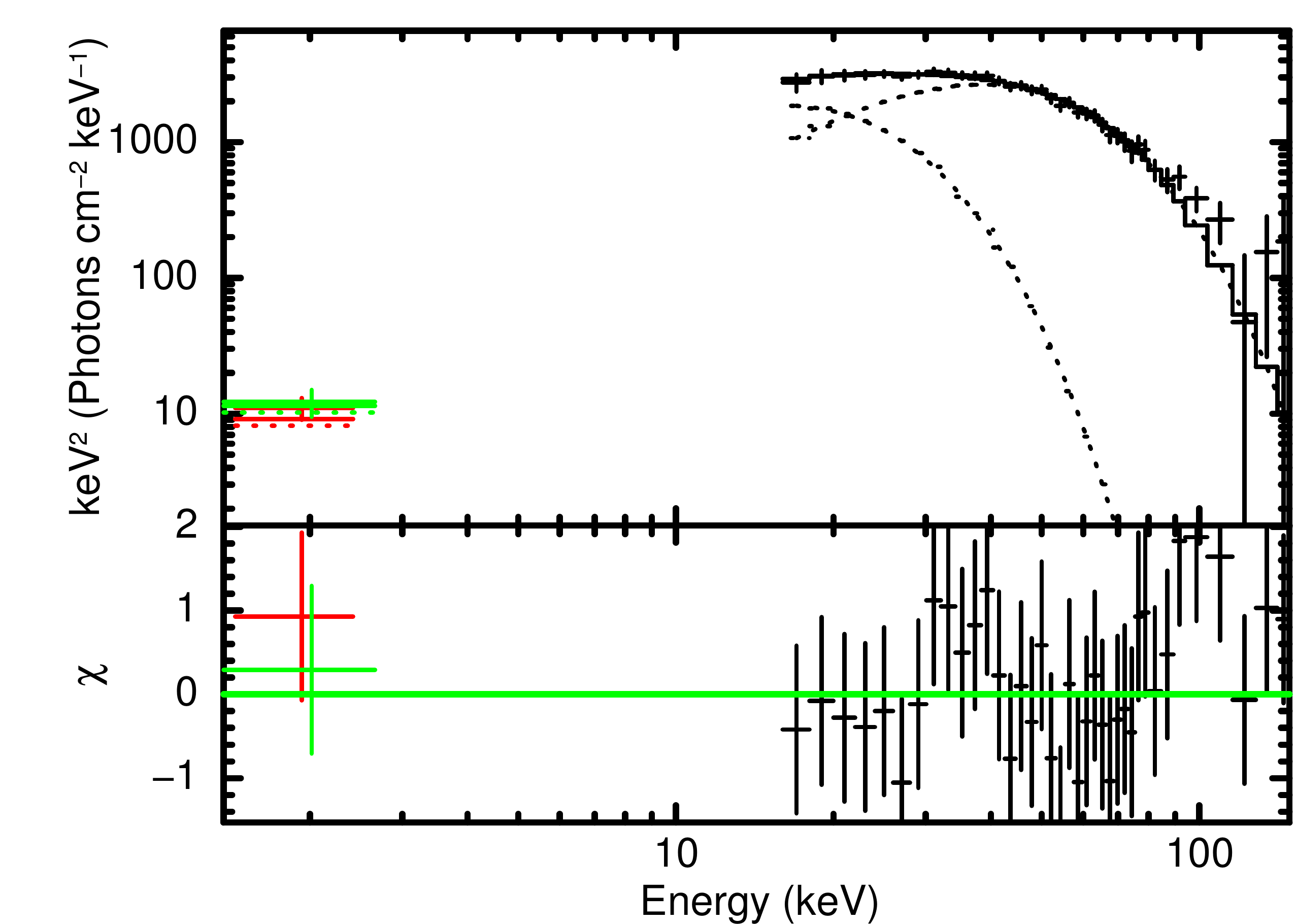}
    \caption{\bat\ (black) and RGS (green and red)  fluence spectra of the burst. Top panel: best fit  with the sum of two blackbody models. 
 Bottom panel: residuals of the best fit.   }
        \label{burst_spectrum}
\end{figure}

\subsection{Burst properties}
\label{sec_burst}

The initial part of the  brightest  burst could not be studied with  EPIC   because the saturation effect due to the high count rate caused a severe loss of photons. 
Therefore, to derive the burst fluence, we  used the  \bat\ and the  RGS data which  were not affected by saturation.

The BAT data show that the burst started at T$_0$=20:00:39.47 UT of  2009 February 3. Its light curve,  in the energy range 15--150 keV  (Fig.~\ref{bat_lc}), was characterized by a bright peak lasting  $\sim$0.1 s,
followed by fainter emission lasting about 0.2 s. 
The burst was also clearly visible in the RGS data, with a start time consistent with the above value, considering the limited time  resolution of the two cameras (read-out times of $\sim$5 s and $\sim$10 s).
 
To estimate the burst fluence, we extracted the BAT and RGS1 and RGS2 spectra of the burst events only, for time intervals of $\sim30$ ms and 10 s, respectively. Since at these time scales the persistent emission contribution is negligible, we studied the burst spectrum in fluence units by assigning an exposure time of 1 s to both RGS and BAT spectra.
Despite the low  RGS count statistics allowed us to use only a single energy bin for each camera (0.3--2 keV), these measurements, when fitted together with the BAT spectrum, could constrain the  burst spectral shape at low energies.
We fitted the spectra simultaneously in the   range 0.3--150 keV,  with the  column density fixed at the value of $4.2\times10^{22}$ cm$^{-2}$ derived for the persistent emission. 
The best-fit  was obtained with the sum of two blackbody models, as found for the spectra of the short bursts of other magnetars in similar energy ranges \citep{feroci04,israel08short}. Our best fit gave  temperatures kT$_1$ = $4.1\pm0.4$ keV and kT$_2$ = $9.8\pm0.4$ keV and   emitting radii R$_1=(14.8\pm3.0)\, d_{4\text{kpc}}$ km and R$_2=(3.1\pm0.3)\, d_{4\text{kpc}}$ km ($\chi^2/dof = 44.74/56$; Fig.~\ref{burst_spectrum}). The absorbed (unabsorbed)   fluence in the   range 0.3--150 keV was  $9.7\times10^{-6}$ erg cm$^{-2}$ ($9.9\times10^{-6}$ erg cm$^{-2}$), while in the 4--10 keV  range it was $1.1\times10^{-6}$ erg cm$^{-2}$ ($1.2\times10^{-6}$ erg cm$^{-2}$).

\subsection{Timing and spatial analysis of the burst tail}
\label{tail_spatial}

\begin{figure}
	\hspace{-0.1cm}\includegraphics[width=8.8cm]{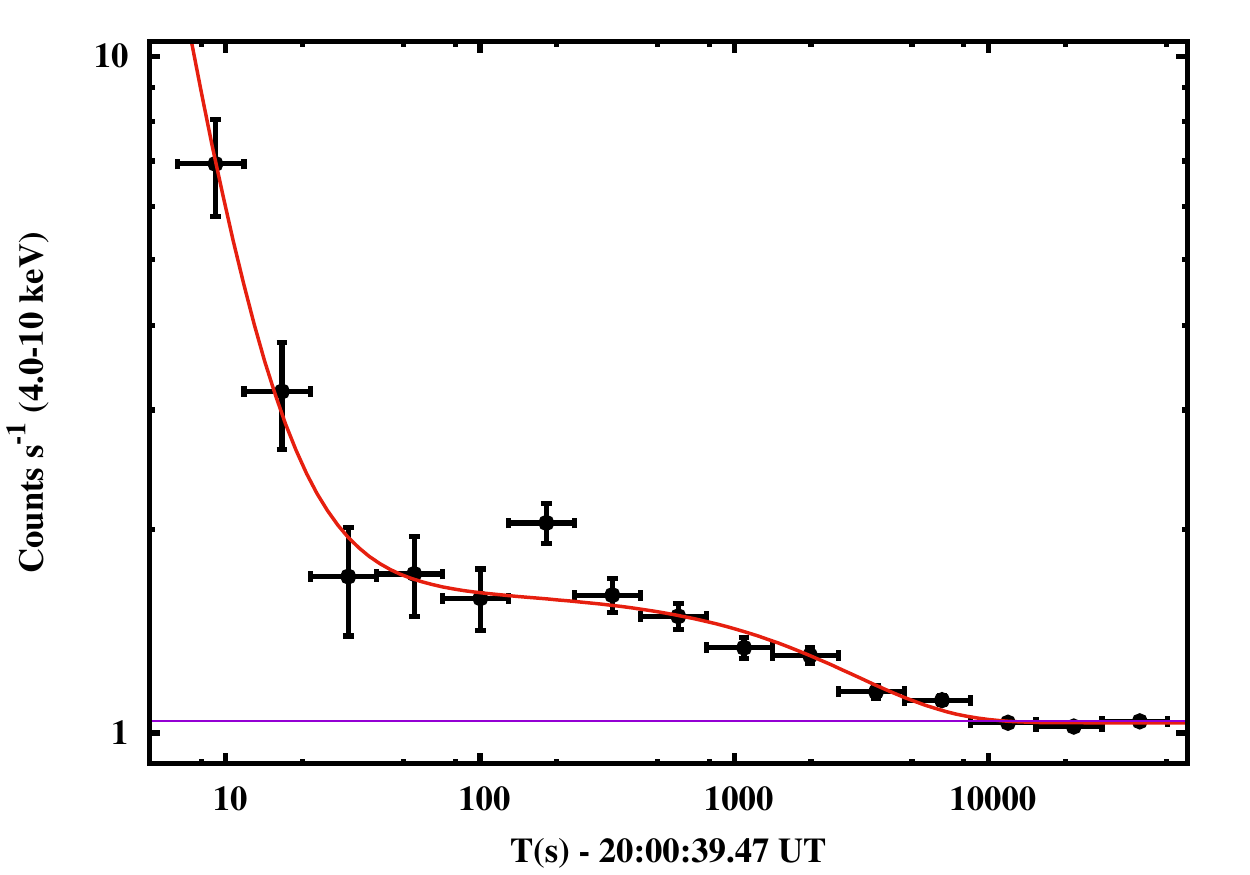}
    \caption{ 4--10 keV EPIC-pn   lightcurve  of the tail after the main burst.  The horizontal line indicates the level of the persistent emission. The red line is the fit with a power-law and an exponential as described in the text.}
        \label{fit_rebin}
\end{figure}

The bright burst described in the previous subsection was followed by an extended tail of X-ray emission
(Fig.~\ref{fit_rebin}). The initial part of the tail showed a steep  power law decay with  $F(t-T_0) \propto (t-T_0)^{-(2.3\pm0.9)}$ cts s$^{-1}$, while the following part, with the exception of a rebrightening at $t-T_0 \sim 200$ s,  can be described by an exponential function, $F(t-T_0) =   (0.47\pm0.09)e^{-(t-T_0)/(2496\pm473)}$  cts s$^{-1}$.

\begin{figure}
	\hspace{-0.7cm} \subfigure{\includegraphics[width=25cm]{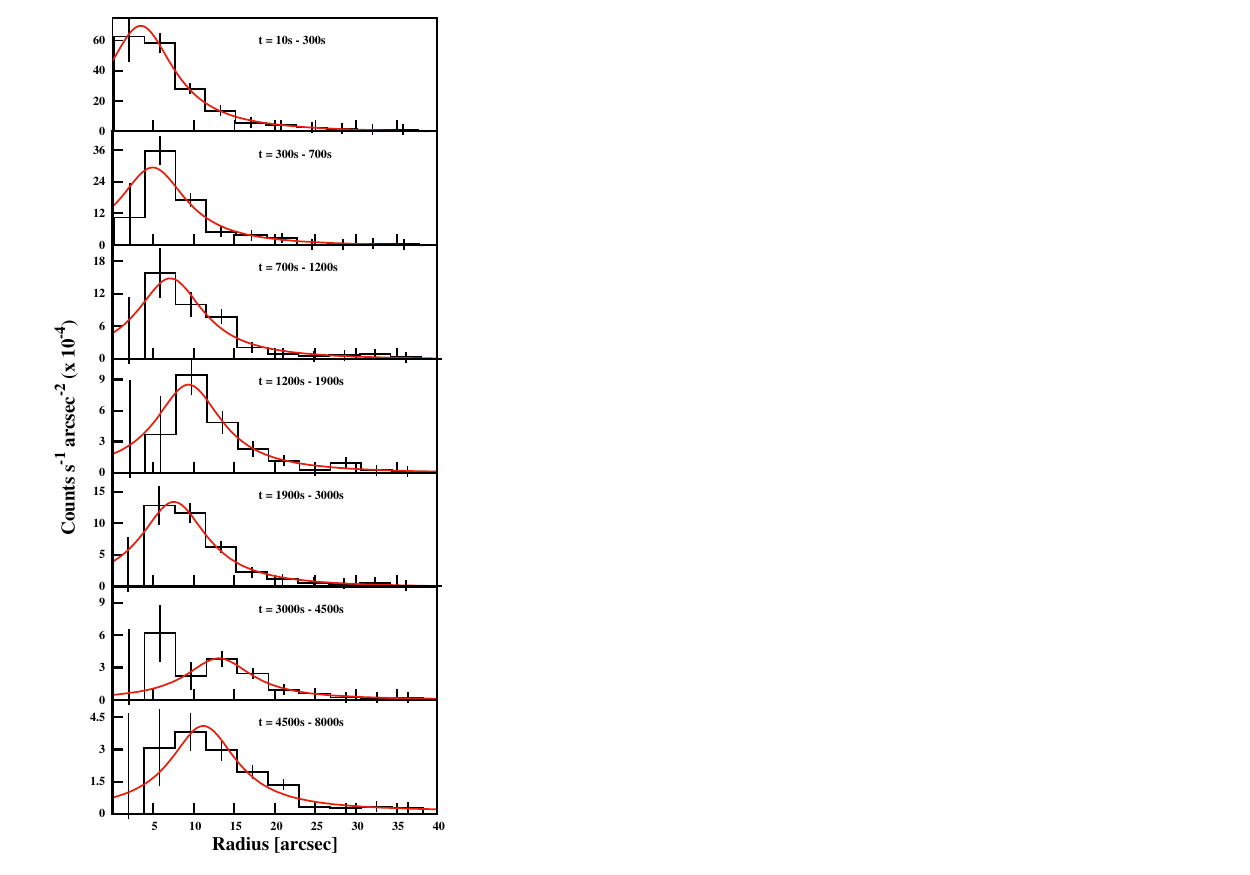}}
\vspace{-0.5cm}
    \caption{Radial profiles of summed EPIC-pn and MOS data at different time intervals starting from the burst, after subtracting the profile of the persistent emission. The solid line is the best fit of the profile using a constant plus a lorentzian function. The data clearly show a ring expanding and fading with time.}
        \label{Radial_profile_fig}
\end{figure}

To study the angular distribution of the tail emission and its time evolution,  starting from the burst epoch, we built the radial profiles for  several, consecutive, time intervals of different durations from  the summed  pn and MOS images. 
We subtracted from these profiles the radial profile of   the persistent emission,  derived from the  same time interval used for the spectral analysis. 
The resulting net radial profiles, shown in Fig.~\ref{Radial_profile_fig}, suggest the presence of excess emission  
at an angular distance which increases with time.  
The fit of the radial profiles with a King function (as expected for the EPIC PSF) was not acceptable ($\chi^2_{\nu} >2$)  in all but the first two time intervals, clearly indicating   time variability of the   profiles,   possiby due to the presence of dust-scattering expanding rings. Note that these deviations from a King profile cannot be attributed to pile-up, since the importance of  this effect increases with the source count-rate and it should therefore affect the first time intervals rather than the ones found to be inconsistent with the PSF.
We fitted the net radial profiles with  a Lorentzian (see  \citealt{tiengo10}), plus a constant to account for variable diffuse emission on large spatial scales. 
We fixed  the width of the  Lorentzian to  10$''$,   corresponding to the value expected for a geometrically-thin ring profile broadened by  the EPIC PSF. With this model we found a significant improvement in the quality of the fits for all the   radial profiles ($\chi^2_{\nu}\sim1$; Fig.~\ref{Radial_profile_fig}). 

\begin{figure*}
	\subfigure{\includegraphics[height=8.8cm,angle=270,]{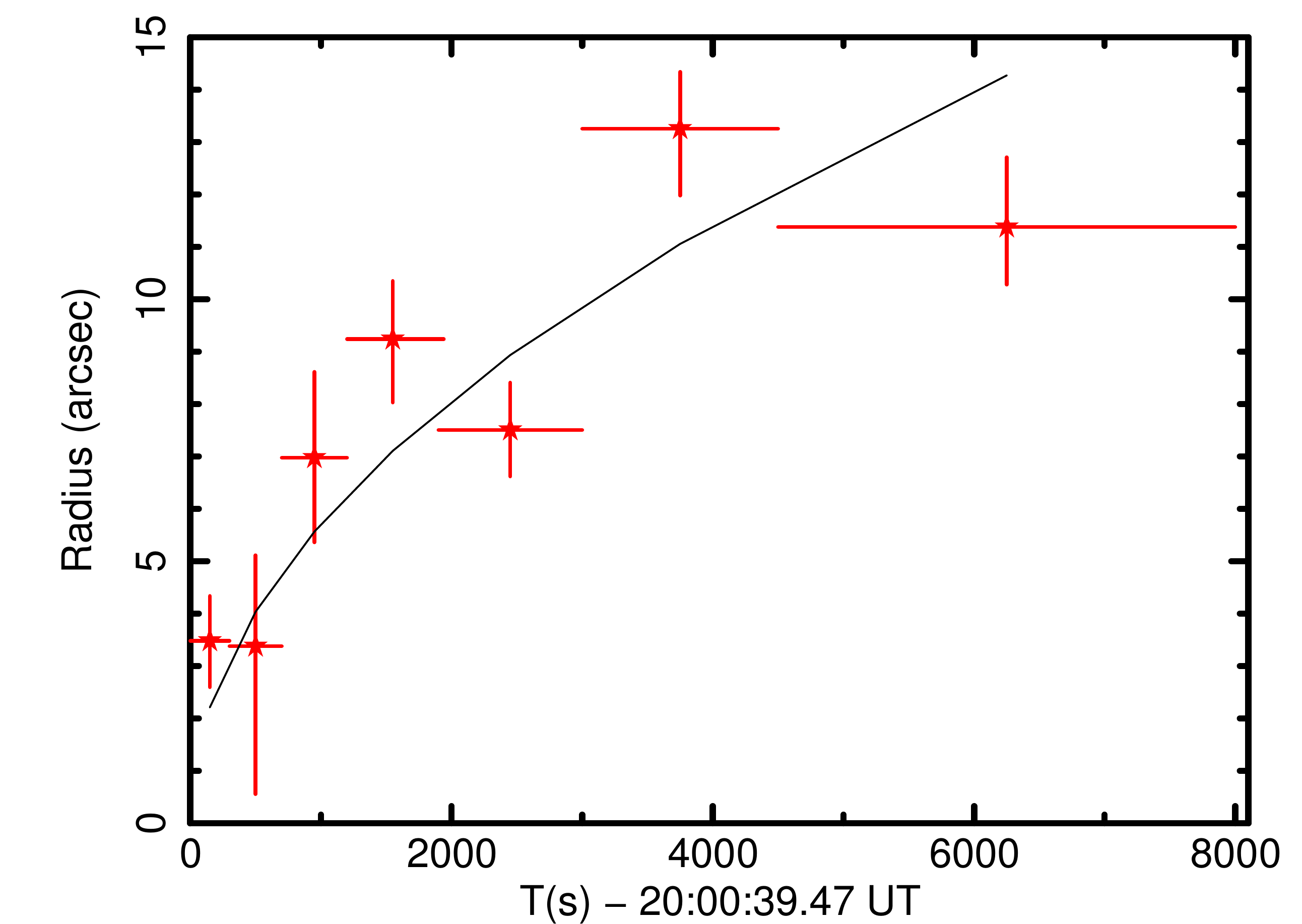}}
	\subfigure{\includegraphics[height=8.8cm,angle=270,]{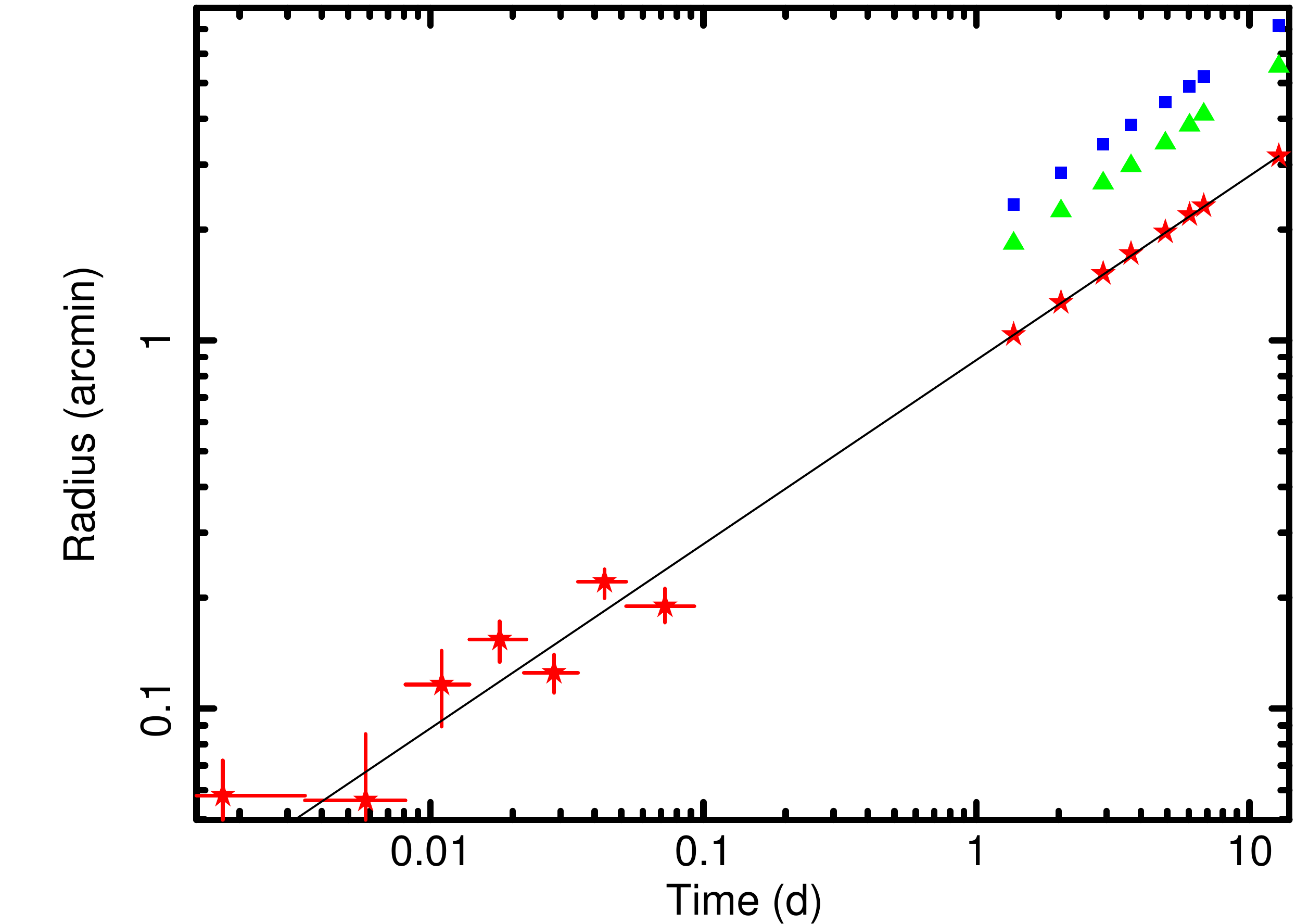}}
    \caption{Left: angular size of the inner ring as a function of time; each point (red) was obtained by fitting with a constant plus a lorentzian model the radial profile obtained in consecutive bin times (with varying length) starting from the burst; the solid line is the best fit $\theta(t) = (0.884\pm 0.045) \cdot t^{0.5}$. Right: joint fit with the points taken from \citet{tiengo10}. The red stars, green triangles, and blue squares indicate the angular radii of the three rings caused by the burst of January 22.   Error bars are at 1$\sigma$ in both figures.}
        \label{radi_prof}
\end{figure*}

The  evolution of the centroid of the Lorentzian component as a function of  time is shown in the left panel of  Fig.~\ref{radi_prof}. 
We fitted the centroid values with the model expected to describe the angular expansion, $\theta(t) = K(t-T_0)^{0.5}$,
and we found $K=0.884\pm0.045$ arcmin day$^{-0.5}$ (error at $1\sigma$). This value is fully consistent with the value K=0.8845$\pm$0.0008 reported in \citet{tiengo10} for the  farthest\footnote{The farthest dust layer is responsible for the smallest of the three rings of  \citet{tiengo10} (``inner'' ring, in the following).} of the three dust layers present between us and \src.
We could better estimate $K$  by fitting the new data points together with those reported in \citet{tiengo10} (right panel of Fig.~\ref{radi_prof}). The best fit gave $\chi^2/dof=25.31/15$ and $K=0.8847 \pm 0.0008$ arcmin day$^{-0.5}$. If instead we  fit the new points with those associated to the other two dust layers,   we obtain significantly poorer fits (reduced $\chi^2 >2$).  
Finally, we also searched for the rings produced by the other two dust layers in the radial profiles by fitting them with two additional Lorentzian components, but they were too weak to be detected.

\subsection{Dust scattering modelling of the tail light curve}

The  above results indicate that at least part of the tail emission is consistent with a dust scattering ring which was expanding at the same rate measured by \citet{tiengo10} for the inner ring produced by  the strong burst of January 22.

As a further check,  we  computed the  light curve expected for the scattered emission  in the 4--10 keV band, where the tail is more evident.
We  adopted the  grain model that gave the best fit to the data of \citet{tiengo10},  i.e. the \textit{BARE-GR-B}   
model\footnote{This model includes polycyclic aromatic hydrocarbons (PHA), silicate grains and  bare graphite grains with the abundances of B-type stars.} of \citet{zubko04}, and the corresponding distances   for \src\ (3.9 kpc) and for the three dust layers  (3.4, 2.6 and 2.2 kpc). 

We used the spectral distribution and fluence of the  burst derived in section \ref{sec_burst} and included the effect of the three dust layers with column densities fixed at $1\times10^{22}$, $0.24\times10^{22}$ and $0.27\times10^{22}$ cm$^{-2}$ \citep{tiengo10}. 
We also assumed that the dust layers are geometrically thin and that the burst was instantaneous.

\begin{figure}
	\subfigure{\includegraphics[width=8.7cm]{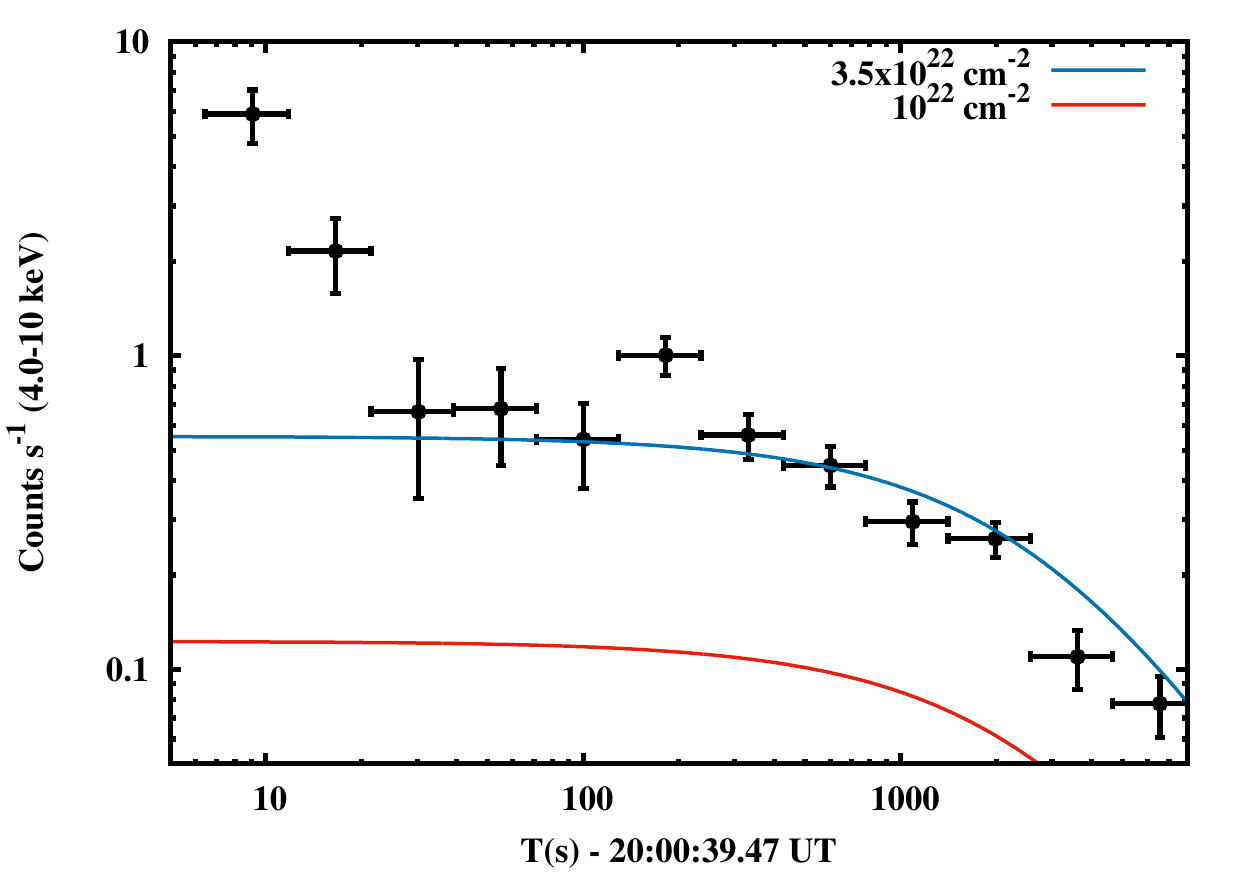}}
    \caption{Comparison between the  4--10 keV  EPIC-pn lightcurve (black points, persistent emission subtracted), and the simulated lightcurves of dust-scattering for the \textit{BARE-GR-B} grains model \citep{zubko04}. The column density of the farthest dust layer is $10^{22}$ cm$^{-2}$ (red line) and $3.5\times10^{22}$ cm$^{-2}$ (blue   line).}
        \label{lc_comp}
\end{figure}

In Fig.~\ref{lc_comp} we compare the computed lightcurve (red line) with the   observed data. 
Clearly, we  cannot reproduce the   data with the above column density values.
Instead, we found that  a $\sim$3.5 times higher column density in the three dust layers is required  to describe the  tail lightcurve after  the first $\sim20$ s (blue line in Fig.~\ref{lc_comp}).

\subsection{Spectral analysis of the tail}
\label{tail_spec}

The results described above indicate that the angular distribution and the light curve of the X-ray tail, at least for times greater than $\sim T_0$+20 s, are consistent with those  expected for dust scattering.
Here we investigate whether also the spectral properties of the tail emission are consistent with this hypothesis.
 
We   created an  {\sc xspec}  table model  with the spectra expected for the scattered emission. We computed it for a range of  distances  of \src\  (from 0.1 to 10 kpc) and using again the \textit{BARE-GR-B}  dust model.
For each value of the source distance, the relative distances of the three dust layers are determined by the measured rates of expansion of the rings   \citep{tiengo10}. We assumed that the optical depths of the two closer dust layers are 24\% and 27\% of that of the layer responsible for the inner ring. We fixed the burst spectrum and its fluence as done above for the light curve computation.
Therefore, the table model has only two parameters: source distance and  column density  of the farthest dust layer, N$_{\text{H1}}$.

We fitted  the pn  spectrum of the tail extracted in the time interval from $T_0+20$ s to  $T_0+4$ ks. The limit of 4 ks was chosen because,  for longer exposure times, the signal-to-noise ratio of the spectrum is too low. 
The contribution of the persistent emission was included in the fit, with  parameters fixed at the best-fit blackbody plus powerlaw model  described above.

\begin{figure}
	\hspace{-0.5cm}\includegraphics[width=6.2cm,angle=270]{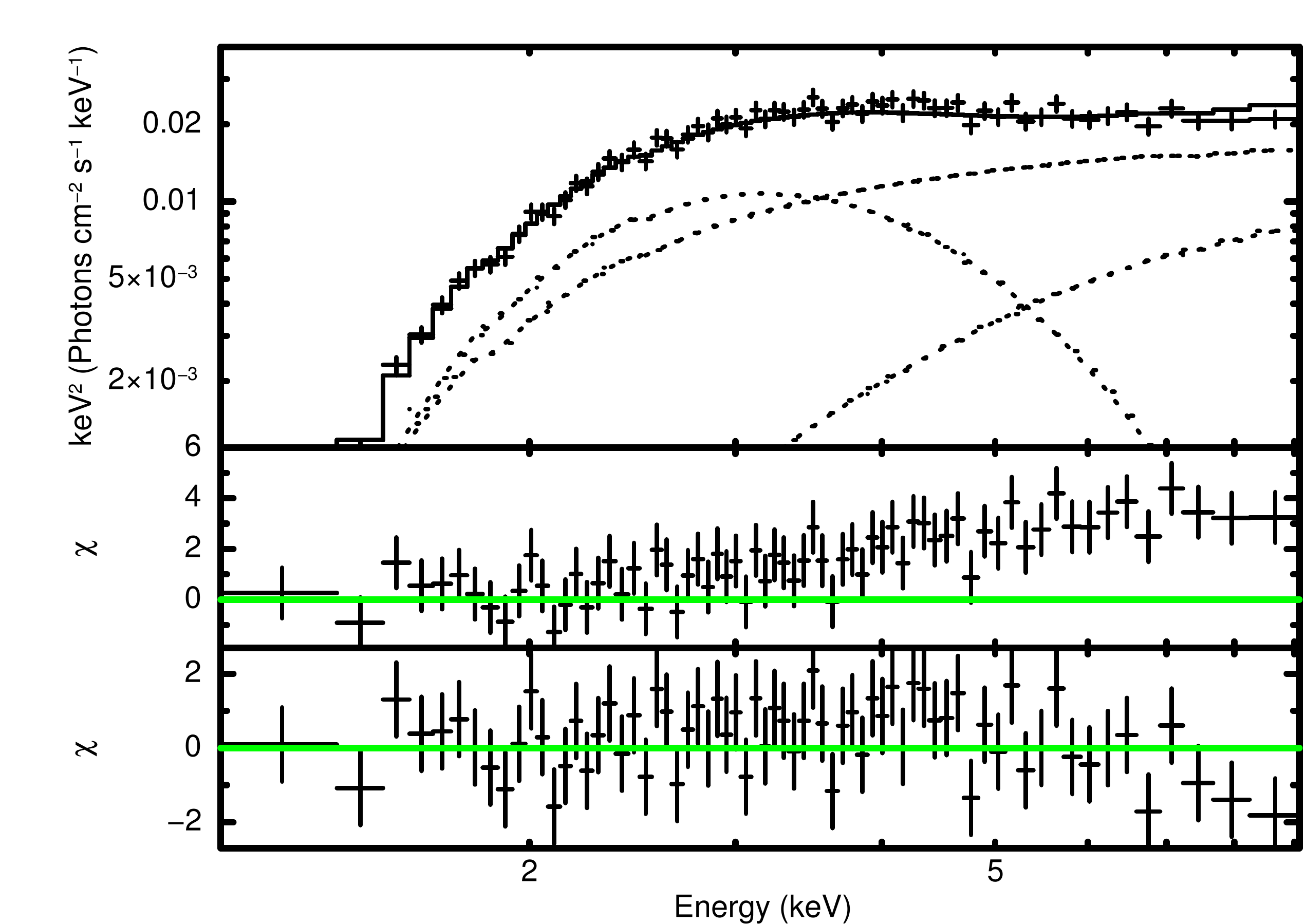}
    \caption{Top panel: spectrum of the tail integrated for 4 ks after the burst. The solid line is the best-fit model, while the dashed lines represent the blackbody,  powerlaw, and  dust-scattering components.
    Center panel: residuals of the fit when adopting only the best fit model of the persistent emission. Bottom panel: residuals of the fit when adding the dust-scattering component to the persistent emission.}
        \label{epoch2}
\end{figure}

Since the source distance was poorly constrained, we fixed it at 3.9 kpc  and found a very good fit ($\chi^2$/dof = 62.79/67) with N$_{\text{H1}}=(4.0\pm0.4)\times10^{22}$ cm$^{-2}$   (see Table~\ref{spectral_model} and Fig.~\ref{epoch2}). 
The total  column density in the three dust layers,  N$_{\text{H1}}\times(1+0.24+0.27) = (6.0\pm0.6) \times 10^{22}$ cm$^{-2}$  is about a 40\% larger than that derived from the fit of the persistent emission (N$_{\text{H1}}=4.2\times10^{22}$ cm$^{-2}$). Fixing  N$_{\text{H1}}=2.78\times10^{22}$ cm$^{-2}$, to avoid exceeding the total absorption of the persistent emission, requires an additional blackbody to fit the spectrum. Its parameters are kT $=1.2$ keV and emitting radius of R $=167_{-167}^{+95}$ m. Such  a thermal component can be attributed to the emission from a small region on the NS surface that has been heated by the energy released in the burst. Although the current data yield relatively large errors on the best fit parameters, we note that by  completely neglecting the dust scattering, one would obtain different results. In fact, if we assume that there is no dust scattering contribution in the tail, it is still possible to obtain a good fit  ($\chi^2/dof = 49.97/66$) by adding to the persistent emission  a  blackbody with a slightly higher temperature  kT =$1.8_{-0.2}^{+0.3}$ keV, emitting radius   R $=135_{-30}^{+35}$ m,  and   flux of $(7.1\pm1.0)\times10^{-12}$ erg cm$^{-2}$ s$^{-1}$ (4--10 keV unabsorbed, see Table~\ref{spectral_model}).

\begin{table*}
          \caption{Spectral results for the burst tail} 
      \label{spectral_model}
\begin{tabular}{lcccccl}
\hline
&  \multicolumn{2}{c}{{\sc blackbody}} & \multicolumn{2}{c}{{\sc dust scattering}}  & \\
&  \multicolumn{1}{c}{kT} &  \multicolumn{1}{c}{Radius$^a$} &   \multicolumn{1}{c}{N$_{\text{H1}}^b$} &    \multicolumn{1}{c}{Flux$^c$} & $\chi^2$/dof \\
&  \multicolumn{1}{c}{(keV)} & \multicolumn{1}{c}{(m)}  &  \multicolumn{1}{c}{($10^{22}$ cm$^{-2}$)} &     \multicolumn{1}{c}{( 10$^{-12}$ erg cm$^2$ s$^{-1}$) } &  \\
\hline
\\
{\it Only NS}     &  $1.8^{+0.3}_{-0.2}$  & $135^{+35}_{-30}$  & -                            &                                - & 49.97/66\\
\\
{\it Only dust } &                                  - &    -                     & 4.0$\pm$0.4 &   9.9$\pm$1.1 & 62.79/67\\
\\
{\it NS+dust }  &   $1.2^{+0.4}_{-0.2}$ &  $167^{+35}_{-167}$   & 2.78 (fixed)             &   5.7$\pm$0.7  & 50.90/66\\
\hline

\end{tabular}
\begin{flushleft}
Notes: The model of the persistent emission has been included in all the fits with fixed parameters (see text). The total absorption and the source distance have been fixed at N$_{\text{H}}=4.2\times10^{22}$ cm$^{-2}$  and  d=3.9 kpc, respectively.   Errors   at $90\%$  confidence level. \\
$^a$ Blackbody emitting radius for d=3.9 kpc.\\
$^b$ Column density of farthest dust layer. \\
$^c$ Average flux in the 0.3--10 keV range (unabsorbed). 
 \end{flushleft}
\end{table*}

\section{Discussion}
\label{discussion}

 In this work we studied a long-lasting X-ray tail  emitted from the magnetar \src\ after  a bright short burst on 2009 February 3. 
The burst consisted of a bright peak lasting about 0.1 s, with  fluence,  derived with a joint spectral analysis of \bat\ and \xmm /RGS data,  of  $\sim10^{-5}$ erg cm$^{-2}$ (0.3-150 keV).
The  X-ray tail was characterized by a steep initial decay, with a duration of about 20 s, followed by a slower decrease, lasting at least 8 ks and with a fluence of $\sim4\times10^{-8}$ erg cm$^{-2}$.
We found evidence for an angular expansion of the X-ray emission during the tail and, 
thanks to the previous knowledge of the properties of the  dust clouds in the direction of \src\  \citep{tiengo10},  we  showed that the part of the tail starting  $\sim$20--30 s after the burst  can be entirely explained by  dust-scattering.

\begin{table*}
  \begin{center}
     \caption{Bursts and flares with extended X-ray tails in magnetars.} 
      \label{burst_prop}
\begin{tabular}{lcccccl}
\hline
Source &  Burst   &  F$_{BURST}^a$      & Tail duration  &  F$_{TAIL}^a$       &  Ref.$^b$ & Notes\\
(distance)    &   date         &  (erg cm$^{-2}$) & (ks)             & (erg cm$^{-2}$)                 & &    \\
\hline
  SGR 1900+14      & 1998/08/27 &  $>$8$\cdot10^{-3}$      & 0.4             &  1.3$\cdot10^{-2}$    & M99 & Giant flare \\
  (14 kpc)  			 &1998/08/29 &   1.4$\cdot10^{-4}$          & $\sim$8     &  6.7$\cdot10^{-7}$    & I01,L03  &\\
                           	 &2001/04/28 &   1.8$\cdot10^{-4}$        & $\sim$4       & 2.8$\cdot10^{-7}$     & L03 & \\
\hline
 SGR 1806--20    & 2004/06/22 &  1.8$\cdot10^{-5}$         & $\sim$0.9    & 3$\cdot10^{-8}$        & G11 & \\
 (8.7 kpc)             & 2004/10/17 &  1.3$\cdot10^{-4}$         & $>$1.3     &  7$\cdot10^{-7}$        & G11  &\\
                           & 2004/12/27 &   1.4                                 &  0.4              &  2.5$\cdot10^{-2}$    & F07 &Giant flare  \\ 
\hline
 SGR 0526--66   &     1979/03/05 &  5$\cdot10^{-4}$         & 0.2              &  4$\cdot10^{-3}$         & M99 & Giant flare\\
(55 kpc) &          &    &   &    &   &   \\
\hline
1E 1547--5408 &   2009/01/22 &    $>$6$\cdot10^{-5}$ & 0.00785         &  6.9$\cdot10^{-4}$      & M09,T10 & Tail at E$>$80 keV \\
   	(3.9 kpc)	  & 2009/01/22 &    2.65$\cdot10^{-7}$      & 0.017            &   5.97$\cdot10^{-8}$   & M15 &   \\
 				  & 2009/02/03 &    9.9$\cdot10^{-6}$    & $\sim$8            & 3.7$\cdot10^{-8}$       &   This work  & Total tail emission\\
 			       &                   &                                         & $\sim$0.030    & 1.8$\cdot10^{-9}$       &    This work  & Corrected for dust scattering\\
 		     	       & 2009/02/06 &   1.29$\cdot10^{-9}$      & 3.534              &   2.9$\cdot10^{-7}$     & M15  &\\
	 			 & 2009/03/30 &    4.3$\cdot10^{-7}$    & 0.624                 &   3.1$\cdot10^{-8}$     & M15,K12 &  \\
 				 & 2010/01/11 &   1.1$\cdot10^{-8}$     & 0.773                 &   4.9$\cdot10^{-8}$     & M15,K12 & \\
\hline
 4U 0142+61   &  2006/06/25  &    2$\cdot10^{-9}$    & $>$0.463           &  2.86$\cdot10^{-8}$     & C16 & \\
 	(3.5 kpc)    & 2007/02/07  &  4.17$\cdot10^{-8}$   & $>1.6$              &  1.58$\cdot10^{-7}$      & C16 &\\
		            &  2015/02/28 &  7.63$\cdot10^{-8}$    &    0.3                & $>$9.5$\cdot10^{-8}$   & G16 &  \\
\hline
 XTE J1810--197 &  2004/02/16 & $>$6.5$\cdot10^{-9}$ & $>0.575$ &    $>$3.9$\cdot10^{-8}$  & W05  &\\
(5 kpc)   & &   &   &          &  &\\
\hline
\end{tabular}
\end{center}
\begin{flushleft}
$^{a}$ Unabsorbed fluence in the range 0.3--150 keV. \\
$^{b}$   
C16: \citet{chakraborty16};
F07: \citet{frederiks07};
G11: \citet{gogus11b};
G16: \citet{gogus16b};
I01: \citet{ibrahim01};
K12: \citet{kuiper12};
L03: \citet{lenters03};
M09: \citet{mereghetti09};
M99: \citet{mca99_2};
M15: \citet{mush15};
T10: \citet{tiengo10};
W05: \citet{woods05};

 \end{flushleft}
\end{table*}

The total column density derived from our fit of the X-ray tail with a dust scattering model ($N_{\rm H} = 6\times 10^{22}$ cm$^{-2}$) is slightly larger than that obtained in the spectral analysis of the persistent emission of \src\  ($N_{\rm H} = 4.2\times 10^{22}$ cm$^{-2}$). Such an apparent discrepancy is   likely related to the fact that the gas-to-dust ratio is not uniform in the Galaxy, while the normalization  of the adopted dust model (number of grains per H atom) is based on an average value. 
In particular, the comparison between the $N_{\rm H}$ derived from X-ray spectra and the reddening in stars behind nearby molecular clouds indicates, as in our results,  an excess of dust with respect to gas in dense clouds \citep[see, e.g.,][]{vuong03,hasenberger16}.
On the other hand, if also unscattered emission from \src\ contributes to the tail, a lower   column density of dust is required (see Table~\ref{spectral_model}). Note, however, that such a contribution must be small because, although the tail spectrum is compatible with thermal emission from a fraction of the NS surface, the spatial analysis described in Sect.~\ref{tail_spatial} shows that most of the tail emission is produced by dust scattering.

The steep  part of the X-ray tail  at $t<T_0+$20 s cannot be reproduced by the dust scattering model used to describe it at later times. In principle, it could be explained by invoking the presence of a dense dust layer very close  (less than a few pc) to the source, possibly related to the supernova remnant associated with \src\ \citep{gelfand07}. However, it is more likely that the first part of the X-ray tail is dominated by unscattered emission, which comes directly  from \src . This emission could originate, for example,  in a region of the NS surface which has been heated  by the burst or in a trapped fireball in the star magnetosphere \citep{thompson95}.
 Unfortunately, the small number of counts did not allow us to confirm this with a significant detection of pulsations in the  first 20--30 s.

\begin{figure}
	\includegraphics[width=11.75cm]{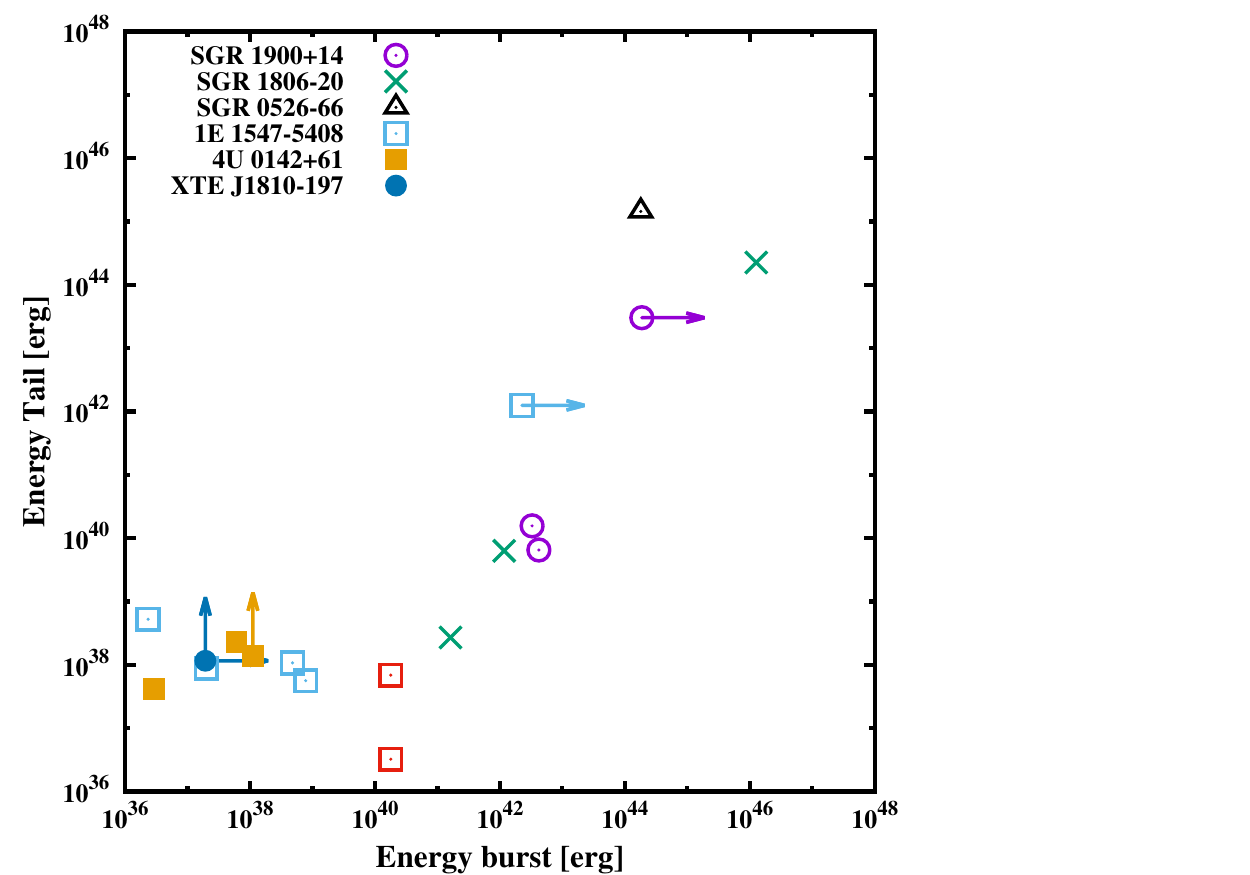}
    \caption{ Energy of the tails versus the energy of their corresponding bursts (see Table~\ref{burst_prop}). 
    The red squares refer to the burst of 2009 February 3, uncorrected and corrected for the dust scattering contribution to the observed tail fluence.}
        \label{brsts}
\end{figure}

To put our results in the context of other similar events, we compiled in Table~\ref{burst_prop} the properties of several bursts followed by  X-ray tails detected in \src\ and in other magnetars. We included, for comparison, also the three giant flares from SGR 0526--66,  SGR 1806--20, and SGR 1900$+$14, although they involved a much larger energy budget \citep[see][and references therein]{mereghetti08}. Based on the best available spectral and timing information reported in the literature, we estimated for each event  the fluence of the tail and that of the associated initial burst and, using the distances given in Table~\ref{burst_prop}, we derived the energies plotted in Fig.~\ref{brsts}. Note that some of these estimates have relatively large uncertainties due to the poorly constrained spectra and/or instrumental saturation effects  (in some cases, only lower limits could be established).  

For the burst of \src\ we plotted two values  in Fig.~\ref{brsts}:  one in which we consider the whole energy of the tail and one in which we exclude the part attributed to dust scattering. For the latter we only considered the fluence of the first part of the tail, estimated by its EPIC-pn spectrum.
The fluence of the short burst corresponds to a total energy release of   $\sim2\times10^{40} \, d^2_{4\text{kpc}}$  erg.  Therefore, this was  a relatively strong burst, although not as energetic as the brighter events seen from this source on 2009 January 22 \citep{mereghetti09,savchenko2010}. 
 
From Fig.~\ref{brsts}, it is clear that magnetar bursts span a large range of intensities, with several events with ``intermediate''  energy outputs, which bridge the ``standard'' bursts with the much rarer giant flares.
Although there is an overall correlation between the burst and tail energetics, especially when also the giant and intermediate flares are considered, the data show a large dispersion. The energy in the tail can  be either a fraction of that of the burst or exceed it.  The latter case seems to require an additional energy supply released on a longer timescale after the burst,  but the analysis  of \src\  reported here shows that, in some cases, the tail energy might be overestimated if assuming an intrinsic NS emission only. 
 
\subsection{Implications for the burst of 2009 January 22}

\citet{tiengo10} studied three bright  dust scattering rings around \src .
Their angular expansion rate   measured with \xmm\ and {\it Swift},  clearly indicates that they were produced by  one (or possibly more) of the very luminous bursts seen by {\it INTEGRAL} in the time interval from 6:43 to 6:51 UT of 2009 January 22  \citep{mereghetti09}.  Unfortunately, the fluence of these bursts in the soft X-ray band (0.5-10 keV)  was not measured directly, since they were observed only in the hard X-ray range and, furthermore, the two brightest ones saturated the instruments.  
For this reason,   \citet{tiengo10}  could obtain only  a combined information of the two factors which determine the flux of the scattered radiation, i.e.  the dust optical depth and the fluence of the burst. 

This degeneracy can now  be removed because, for the burst of February 3 analyzed in this work, both the unscattered  and scattered X-ray fluences have been measured: if the tail after T$_0$+20 s is entirely due to dust scattering, a column density $N_{\rm H1} = 4\times 10^{22}$ cm$^{-2}$ is required for the farthest of the three dust clouds. This is four times larger than the value {\it assumed} by  \citet{tiengo10} and  implies  that  the January 22 burst emitted  $2.5\times10^{43} \, d^2_{4\text{kpc}}$ erg ($4.75\times10^{43} \, d^2_{4\text{kpc}}$ erg)   in the 1-100 keV range, for a thermal bremsstrahlung spectrum with temperature kT = 30 keV (100 keV).    These values are a factor of 4 smaller than those reported in \citet{tiengo10}.

\section{Conclusions}
\label{conclusions}

We have shown that most of the long lasting X-ray emission detected after a   burst of the magnetar \src\ can be explained as radiation scattered by   interstellar dust.  In fact, based on  previous observations of expanding rings after a   brighter event, there is solid evidence for the  presence of a significant amount of dust concentrated in three clouds along the line of sight of this source.  The effect of this dust on fainter bursts, such as the one analyzed here, is certainly less spectacular, but it cannot be neglected. Indeed, by ignoring it, one would attribute the whole X-ray tail, lasting about 10 ks and with a fluence of $\sim4\times10^{-8}$ erg cm$^{-2}$, to enhanced emission coming directly from \src\ and related to   the burst.

Can a similar scenario apply also to other X-ray tails observed after the bursts of different magnetars?  
Most of these sources are located in the Galactic plane, at distances of several kiloparsecs, and have absoption values similar to, and in a few cases larger than,
that of \src .  Undoubtedly, their radiation is subject to some scattering, as demonstrated by the long-lived halos observed around  
SGR 1806--20  \citep{kaplan02}, SGR 1900+14 \citep{kouveliotou01}, SGR 1833--0832 \citep{esposito11short}, Swift J1834.9--0846 \citep{esposito13}, and  SGR J1935+2154 \citep{israel16}.
On the other hand, some X-ray emission  after bright bursts must come directly from the neutron star. This is evident when pulsations are observed, while the interpretation of unpulsed emission, often showing a hard to soft spectral evolution which suggests a cooling region on the NS, is subject to more uncertainties. Often the quality of the data is insufficient to disentangle the  contribution of dust scattering from that of the intrinsic emission in the observed X-ray tails.
Unfortunately,  a quantitative assessment of this effect,  as we did for \src , can be done only when bright bursts are observed and timely follow-ups with good imaging and sensitivity are carried out to study the X-ray halo evolution.

\section*{Acknowledgements}
This work has been partially supported through financial contribution from  PRIN INAF 2014.
The results are  based on observations obtained with XMM-Newton, an ESA science mission with instruments and contributions directly funded by ESA Member States and NASA, and on data obtained from the HEASARC archive. PE acknowledges funding in the framework of the NWO Vidi award A.2320.0076. EC acknowledges support from a VIDI grant from the Netherlands Organisation for Scientific Research (NWO).
We thank Daniele Vigan\`o who developed part of the software used in this work.

\addcontentsline{toc}{section}{Bibliography}
\bibliographystyle{mn2e}
\bibliography{biblio}

\bsp
\label{lastpage}
\end{document}